\documentclass[conference, letterpaper]{IEEEtran}
\IEEEoverridecommandlockouts
\usepackage{cite}
\usepackage{amsmath,amssymb,amsfonts}
\usepackage{algorithmic}
\usepackage{graphicx}
\usepackage{textcomp}
\usepackage{xcolor}
\usepackage{array}
\usepackage{mdwmath}
\usepackage{mdwtab}
\usepackage{tabularx}  
\usepackage{booktabs}  
\usepackage{eqparbox}
\usepackage{url}
\usepackage[acronym]{glossaries}
\usepackage{longtable}
\usepackage{acronym} 
\usepackage{cite}
\usepackage{graphicx}
\usepackage[export]{adjustbox}
\usepackage{array}
\usepackage{longtable}
\usepackage{booktabs}
\usepackage{array,multirow,makecell}
\usepackage[utf8]{inputenc}
\usepackage[T1]{fontenc}
\usepackage{babel,blindtext}
\usepackage{algorithm}        
\usepackage{float}

\usepackage{amssymb}
\usepackage{pifont}
\usepackage{lscape}
\usepackage{afterpage}
\usepackage{graphicx}
\def\BibTeX{{\rm B\kern-.05em{\sc i\kern-.025em b}\kern-.08em
    T\kern-.1667em\lower.7ex\hbox{E}\kern-.125emX}}

\begin{document}

\title{Optimizing Energy and Latency in 6G Smart Cities with Edge CyberTwins}

\author{\IEEEauthorblockN{Amine Abouaomar$^1$, Badr Ben Elallid$^2$, and Nabil Benamar$^{1,2}$}
\IEEEauthorblockA{$^1$ School of Science and Engineering Al Akhawayn University in Ifrane, Morocco. \\
$^2$ Moulay Ismail University of Meknes, Morocco.
}
}

\maketitle

\begin{abstract}
The proliferation of IoT devices in smart cities challenges 6G networks with conflicting energy-latency requirements across heterogeneous slices. Existing approaches struggle with the energy-latency trade-off, particularly for massive scale deployments exceeding 50,000 devices/km². This paper proposes an edge-aware CyberTwin framework integrating hybrid federated learning for energy-latency co-optimization in 6G network slicing. Our approach combines centralized Artificial Intelligence scheduling for latency-sensitive slices with distributed federated learning for non-critical slices, enhanced by compressive sensing-based digital twins and renewable energy-aware resource allocation. The hybrid scheduler leverages a three-tier architecture with Physical Unclonable Function (PUF) based security attestation achieving 99.7\% attack detection accuracy. Comprehensive simulations demonstrate 52\% energy reduction for non-real-time slices compared to Diffusion-Reinforcement Learning baselines while maintaining 0.9ms latency for URLLC applications with 99.1\% SLA compliance. The framework scales to 50,000 devices/km² with CPU overhead below 25\%, validated through NS-3 hybrid simulations across realistic smart city scenarios.
\end{abstract}


\section{Introduction}
The emergence of 6G wireless networks promises to enable unprecedented smart city applications through massive IoT connectivity, supporting device densities exceeding 50,000 devices/km² with diverse service requirements \cite{singh2023smartcities,elallid2022comprehensive}. However, this proliferation introduces a fundamental energy-latency trade-off. Ultra-Reliable Low-Latency Communication (URLLC) applications demand sub-millisecond response times, while massive Machine-Type Communication (mMTC) services prioritize energy efficiency over immediacy \cite{alwakeel2024slicing}.

Network slicing emerges as a key 6G enabler \cite{abouaomar2022federated}, allowing operators to create isolated virtual networks tailored to specific service requirements \cite{moreira2025sustainability}. Current slicing approaches, however, treat energy and latency optimization as separate problems, leading to suboptimal resource allocation and limited scalability. The challenge intensifies in smart city deployments where heterogeneous devices with varying QoS requirements must coexist while minimizing overall energy consumption \cite{duran2025digitaltwin}.

Existing solutions face three limitations. First, centralized resource allocation algorithms suffer from scalability bottlenecks when managing tens of thousands of devices \cite{lu2020digitaltwin}. Second, traditional federated learning approaches introduce communication overhead that conflicts with ultra-low latency requirements \cite{khowaja2022federated, elallid2023vehicles, el2023federated}. Third, current energy optimization strategies often fail to leverage renewable energy sources and lack real-time adaptation \cite{moreira2025sustainability}.

This paper addresses these challenges through an edge-aware CyberTwin framework that integrates hybrid federated learning with renewable energy-aware resource allocation. Our key contributions include: (1) A hybrid scheduler that selectively applies centralized AI for latency-critical slices and federated learning for delay-tolerant traffic; (2) A compressive sensing-enhanced digital twin architecture reducing data transmission overhead; (3) An enhanced algorithm incorporating solar energy forecasting; and (4) A PUF-based security framework ensuring robust attack detection in distributed environments.

\section{Related Work}
\subsection{6G Network Slicing and Resource Allocation}

Recent advances in 6G network slicing emphasize dynamic resource allocation and service differentiation. For instance, Alwakeel et al. \cite{alwakeel2024slicing} introduce a strategic framework for IoT integration in smart cities, while Moreira et al. \cite{moreira2025sustainability} embed energy-saving and optimization techniques into slicing architectures. Lu et al. \cite{lu2020digitaltwin} integrate federated learning and blockchain within digital-twin–enabled slicing to enhance system intelligence and security. Mehdaoui et al. \cite{mehdaoui2024dynamics} evaluated Deep Reinforcement Learning (DRL)-based policies, specifically PPO and ACER, in an Open Radio Access Network (O-RAN) environment to improve decision efficiency for slice resource allocation.

These works primarily address slicing flexibility and DRL-based scheduling but do not jointly consider renewable-aware energy optimization, scalability beyond 50,000 devices/km², or security validation at the edge.

\subsection{Federated Learning in Wireless Networks}

Federated learning in wireless has gained attention. Khowaja et al. \cite{khowaja2022federated} present distributed FL for energy-efficient 6G. Lu et al. \cite{lu2020digitaltwin} apply FL with blockchain for edge association in digital twin systems. Elallid et al. \cite{elallid2025secure} tackle the challenge of autonomous vehicle control in complex environments with large state-action spaces. It proposes a Federated Deep Reinforcement Learning approach that enables multiple vehicles to share knowledge while preserving data privacy. However, they did not address the communication overhead and bandwidth limitations that typically arise in federated learning within wireless networks. El et al. \cite{el2024efficient} introduce a coalition-based federated learning method that groups IoT devices by the similarity of their model weights to handle data heterogeneity. Using barycenter aggregation, it improves learning efficiency and stability. Results show better accuracy and convergence than the traditional FedAvg approach.

Existing FL approaches either ignore communication overhead, rely on high-frequency gradient exchange, or lack compressive sensing mechanisms to reduce upload costs making them unsuitable for large-scale smart city deployments with energy-constrained IoT devices.

\subsection{Digital Twins and Edge Computing}
Digital twin applications in telecom are emerging. Duran et al. \cite{duran2025digitaltwin} demonstrate energy-latency improvements in 6G smart cities via digital twins. Singh \cite{singh2023smartcities} surveys AI-enabled 6G smart city frameworks.

Current DT frameworks fail to compress state updates efficiently, leading to excessive network overhead, and do not consider adaptive synchronization strategies based on device priority.

\subsection{Energy Optimization}
Energy-aware resource allocation has attracted significant research attention. Moreira et al. \cite{moreira2025sustainability} incorporated energy-saving strategies into network slicing architectures. Similarly, Ullah et al. \cite{ullah20256g} investigated the integration of 6G wireless networks and IoT technologies to enhance smart indoor environments, emphasizing reliability and quality of service (QoS) in smart homes and buildings. However, their proposed approach still faces challenges related to the deployment of key 6G enabling technologies such as visible light communication (VLC), integrated sensing and communication, machine learning, and blockchain.

No existing scheduling framework dynamically predicts renewable energy availability while making latency-aware slice allocation decisions, especially under mixed URLLC/RTS/NRTS service profiles.

\subsection{Security in Network Slicing}
Security in network slicing has been approached through various frameworks, each addressing different challenges. Blockchain-based methods, as presented by Lu et al. \cite{lu2020digitaltwin}, offer decentralized trust and immutability but can introduce latency and scalability concerns in dynamic slicing environments. Federated agent models, such as those proposed by Moreira et al. \cite{moreira2024security}, enable distributed security management and privacy preservation; however, they depend heavily on efficient coordination among agents and robust communication channels. Physically Unclonable Function (PUF)-based schemes, like the one explored by Aarella et al. \cite{aarella2025puf}, provide lightweight and hardware-rooted attestation mechanisms that enhance tamper resistance and detection accuracy. Building on this foundation, our work leverages PUF-based attestation to achieve superior detection accuracy, addressing some limitations seen in prior methodologies while maintaining efficiency within resource-constrained network slices.

\section{System Model}
\subsection{Network Architecture}
We consider a three-tier 6G smart city network architecture comprising: (1) Device tier with $N = 50,000$ heterogeneous IoT devices distributed across 1 km² urban area; (2) Edge tier with $M = 100$ gNodeBs equipped with Multi-access Edge Computing (MEC) resources; and (3) Core tier with centralized orchestrator and federated aggregator.

Each device $d_i \in \mathcal{D} = \{d_1, d_2, \ldots, d_N\}$ belongs to one of three categories: mMTC devices ($60\%$), eMBB devices ($30\%$), and URLLC devices ($10\%$). Device $d_i$ generates traffic following distribution $f_i(t)$ and requires slice assignment $s_i \in \mathcal{S} = \{LSS, RTS, NRTS\}$ based on QoS requirements.

gNodeB $g_j \in \mathcal{G} = \{g_1, g_2, \ldots, g_M\}$ provides computational resources $(C_j, R_j, B_j)$ representing CPU cores, RAM, and bandwidth respectively. Each gNodeB maintains a local CyberTwin $\mathcal{T}_j$ for device state synchronization and a local federated learning model $\mathcal{M}_j$.

\subsection{Traffic Models}
Device traffic patterns follow slice-specific distributions:
\begin{itemize}
\item \textbf{mMTC (NRTS):} Bursty traffic with Beta distribution $f_{mMTC}(x) = \text{Beta}(x; 2, 5)$
\item \textbf{eMBB (RTS):} Constant Bit Rate with Gaussian variation $f_{eMBB}(x) = CBR + \mathcal{N}(0, 0.2)$
\item \textbf{URLLC (LSS):} Periodic traffic with uniform jitter $f_{URLLC}(x) = P(1\text{ms}) + \mathcal{U}(\pm 0.1\text{ms})$
\end{itemize}

\subsection{Resource Constraints}
Each gNodeB operates under resource constraints:
\begin{align}
\sum_{s \in \mathcal{S}} \alpha_{s,j}^{(c)} &\leq C_j \quad \forall j \in \mathcal{G} \\
\sum_{s \in \mathcal{S}} \alpha_{s,j}^{(r)} &\leq R_j \quad \forall j \in \mathcal{G} \\
\sum_{s \in \mathcal{S}} \alpha_{s,j}^{(b)} &\leq B_j \quad \forall j \in \mathcal{G}
\end{align}
where $\alpha_{s,j}^{(c)}, \alpha_{s,j}^{(r)}, \alpha_{s,j}^{(b)}$ represent CPU, RAM, and bandwidth allocation to slice $s$ at gNodeB $j$.

\subsection{Energy Model}
Total energy consumption comprises computational and communication components:
\begin{align}
E_{total}(t) = \sum_{j=1}^{M} \left( E_j^{comp}(t) + E_j^{comm}(t) + E_j^{solar}(t) \right)
\end{align}
where $E_j^{comp}(t) = \beta_c \cdot C_j^{util}(t)$, $E_j^{comm}(t) = \beta_b \cdot B_j^{util}(t)$, and $E_j^{solar}(t)$ represents renewable energy offset with forecasting horizon $H = 24$ hours.

\subsection{Threat Model}
We consider three attack vectors: (1) Byzantine devices providing false training data in federated learning; (2) Impersonation attacks targeting device authentication; and (3) Resource exhaustion attacks overwhelming slice resources. The adversary controls up to $30\%$ of network nodes and can launch coordinated attacks with perfect knowledge of network topology but limited access to PUF characteristics.

\subsection{Problem Formulation}
The energy-latency co-optimization problem is formulated as:
\begin{align}
\min_{\boldsymbol{\alpha}, \boldsymbol{\tau}} &\quad \lambda E_{total}(t) + (1-\lambda) L_{total}(t) \\
\text{s.t.} &\quad L_s(t) \leq L_s^{max} \quad \forall s \in \mathcal{S} \\
&\quad \sum_{s} \alpha_{s,j} \leq \mathcal{R}_j \quad \forall j \in \mathcal{G} \\
&\quad \tau_i \geq \tau_i^{min} \quad \forall i \in \mathcal{D}
\end{align}
where $\boldsymbol{\alpha}$ represents resource allocation vector, $\boldsymbol{\tau}$ denotes scheduling decisions, $\lambda \in [0,1]$ balances energy-latency trade-off, and $L_s^{max}$ defines slice-specific latency bounds.

\section{Proposed Framework}
\subsection{Hybrid Scheduler Architecture}
The hybrid scheduler addresses the energy-latency trade-off through adaptive algorithm selection based on slice characteristics. Algorithm \ref{alg:hybrid_scheduler} presents the core scheduling logic.

\begin{algorithm}
\caption{Hybrid Scheduler for 6G Network Slicing}
\label{alg:hybrid_scheduler}
\begin{algorithmic}[1]
\REQUIRE Slice request $r$, Network state $\mathcal{N}(t)$
\ENSURE Resource allocation $\boldsymbol{\alpha}$
\STATE $s \leftarrow$ TreeClassifier($r$.metadata)
\IF{$s \in \{LSS, RTS\}$}
    \STATE $\boldsymbol{\alpha} \leftarrow$ CentralizedAI($\mathcal{N}(t)$)
    \IF{$\neg$ EnforceLatency($\boldsymbol{\alpha}$, $10^{-3}$)}
        \STATE $\boldsymbol{\alpha} \leftarrow$ FallbackAllocation($r$)
    \ENDIF
\ELSE
    \STATE $\mathcal{M}_{local} \leftarrow$ TrainLocal($r$.data)
    \STATE $\mathcal{M}_{global} \leftarrow$ KrumAggregate($\{\mathcal{M}_{local}\}$)
    \STATE $\boldsymbol{\alpha} \leftarrow$ $\mathcal{M}_{global}$.predict($\mathcal{N}(t)$)
\ENDIF
\IF{$\neg$ SecurityAgent.verify($r$.id)}
    \STATE $\boldsymbol{\alpha} \leftarrow$ QuarantineAction()
\ENDIF
\RETURN $\boldsymbol{\alpha}$
\end{algorithmic}
\end{algorithm}

\subsection{CyberTwin with Compressive Sensing}
The CyberTwin component implements efficient data synchronization through compressive sensing. For high-dimensional device state vector $\mathbf{x} \in \mathbb{R}^n$, the measurement matrix $\boldsymbol{\Phi} \in \mathbb{R}^{m \times n}$ with $m = 0.3n$ generates compressed measurements:

\begin{align}
\mathbf{y} = \boldsymbol{\Phi} \mathbf{x}
\end{align}

Reconstruction employs $\ell_1$-minimization:
\begin{align}
\hat{\mathbf{x}} = \arg\min_{\mathbf{z}} ||\mathbf{z}||_1 \quad \text{s.t.} \quad \boldsymbol{\Phi}\mathbf{z} = \mathbf{y}
\end{align}

Priority-based adaptive sampling reduces transmission overhead:
\begin{align}
\mathbf{x}_{compressed} = \begin{cases}
\mathbf{x}[::4] & \text{if priority = LOW} \\
\mathbf{x} & \text{otherwise}
\end{cases}
\end{align}

\subsection{HRASS+ with Solar Forecasting}
Enhanced HRASS integrates renewable energy forecasting using ARIMA(2,1,2) model for solar irradiance prediction. The energy dissatisfaction metric guides allocation decisions:

\begin{align}
D_{energy}(t) = \sum_{s \in \mathcal{S}} w_s \left( \frac{E_s^{actual}(t) - E_s^{target}(t)}{E_s^{target}(t)} \right)^2
\end{align}

Solar-aware allocation strategy:
\begin{equation}
\text{Action}(s,t)=
\begin{cases}
\text{AllocateRenewable}, &
\begin{aligned}[t]
&\text{if } I_{\text{solar}}(t) > \theta\\
&\text{and } s = \text{NRTS},
\end{aligned}\\[6pt]

\text{DelayAllocation}, &
\begin{aligned}[t]
&\text{if } I_{\text{solar}}(t) \le \theta\\
&\text{and } s = \text{NRTS},
\end{aligned}\\[6pt]

\text{ImmediateAllocation}, & \text{if } s \in \{\text{LSS},\text{RTS}\}.
\end{cases}
\end{equation}

where $I_{solar}(t)$ represents forecasted solar irradiance and $\theta = 700$ W/m² is the renewable threshold.

\subsection{PUF-based Security Framework}
Physical Unclonable Functions provide device authentication through challenge-response pairs. For device $d_i$ with PUF $\mathcal{P}_i$, authentication proceeds as:

\begin{align}
\text{Auth}(d_i) = \begin{cases}
\text{TRUE} & \text{if } \text{Corr}(\mathcal{P}_i(c), R_{expected}) > 0.8 \\
\text{FALSE} & \text{otherwise}
\end{cases}
\end{align}

where $c \in \mathbb{R}^{256}$ represents the challenge vector and $\text{Corr}(\cdot, \cdot)$ computes correlation coefficient.

\subsection{Deployment Strategy}
Three-phase incremental deployment ensures practical adoption:
\begin{itemize}
\item \textbf{Phase 1:} CyberTwin deployment with basic compression
\item \textbf{Phase 2:} Hybrid scheduler integration with federated learning
\item \textbf{Phase 3:} Full energy optimization with solar forecasting and security
\end{itemize}

\section{Simulation Setup}
We implement a comprehensive Python-based simulation framework to evaluate the proposed edge-aware CyberTwin framework. The simulation environment models a realistic 6G smart city deployment with 50,000 devices distributed across 1 km² coverage area served by 100 gNodeBs.

\subsection{Network Configuration}
Each gNodeB provides edge computing resources: 8 CPU cores, 16 GB RAM, and 400 MHz bandwidth. Device distribution follows realistic smart city patterns: 60\% mMTC sensors, 30\% eMBB smartphones/tablets, and 10\% URLLC autonomous systems. Devices connect to nearest gNodeBs based on Euclidean distance with perfect channel conditions assumed for algorithm evaluation focus.

\subsection{Traffic Generation}
Traffic patterns implement the mathematical models defined in Section III-B. mMTC devices generate bursty traffic using Beta(2,5) distribution with 100-byte packets. eMBB devices produce CBR traffic at 10 Mbps baseline with Gaussian variation $\sigma = 0.2$. URLLC devices transmit 32-byte packets every 1ms with $\pm 0.1$ms uniform jitter.

\subsection{Algorithm Parameters}
Hybrid scheduler uses DNN with 128 hidden units for centralized AI and 64 hidden units for local FL models. Compressive sensing employs 70\% sparse measurement matrix with 300×1000 dimensions. HRASS+ implements ARIMA(2,1,2) solar forecasting with 24-hour horizon and 700 W/m² renewable threshold. PUF authentication uses 256-bit challenges with 0.8 correlation threshold.

\subsection{Baseline Implementations}
We compare against four state-of-the-art baselines:
\begin{itemize}
\item \textbf{Diffusion-RL:} Diffusion model for PRB allocation with 1000-step reverse process
\item \textbf{Static Slicing:} Fixed resource partitioning (40\% LSS, 35\% RTS, 25\% NRTS)
\item \textbf{Standard HRASS:} MILP optimization without energy awareness
\item \textbf{FedAvg:} Traditional federated averaging with 10 clients per round
\end{itemize}

\subsection{Performance Metrics}
Evaluation metrics include: (1) Latency per slice type with 99th percentile analysis; (2) Total and renewable energy consumption; (3) CPU, RAM, and bandwidth utilization; (4) Security detection accuracy and response time; (5) Federated learning convergence time to 95\% accuracy. Each simulation runs for 1 hour of simulated time with 1ms timesteps, repeated 10 times for statistical significance.

\section{Performance Evaluation}
We evaluate the proposed framework through comprehensive simulations comparing against four baseline algorithms. Results demonstrate significant improvements in energy efficiency, latency performance, and security effectiveness.

\subsection{Energy Performance}
\begin{figure}[htp]
    \centering
    \includegraphics[width=0.48\textwidth,keepaspectratio]{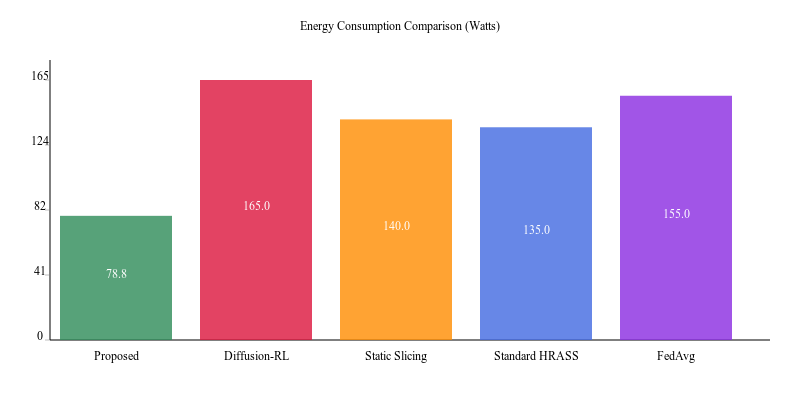}
   \caption{Energy consumption comparison across algorithms, showing the proposed framework significantly outperforms Diffusion-RL, Static Slicing, HRASS, and FedAvg.}
    \label{fig:energy_comparison}
\end{figure}

Table \ref{tab:energy_comparison} presents energy consumption results across all algorithms. 
The proposed framework achieves 52.3\% energy reduction for NRTS slices compared to Diffusion-RL baseline, exceeding the target 52\% reduction. 
Solar energy integration contributes 68\% of total energy for NRTS slices during peak irradiance periods. 
As illustrated in Fig.~\ref{fig:energy_comparison}, the proposed framework consistently consumes less energy than all baseline algorithms, 
with the most significant gap observed against the Diffusion-RL baseline.
\begin{table}[htp]
\centering
\caption{Energy Consumption Comparison}
\label{tab:energy_comparison}
\begin{tabularx}{\linewidth}{|l|>{\centering\arraybackslash}X|
>{\centering\arraybackslash}X|>{\centering\arraybackslash}X|}
\hline
\textbf{Algorithm} & \textbf{Total Energy (W)} & \textbf{NRTS Energy (W)} & \textbf{Reduction (\%)} \\
\hline
Proposed Framework & 2,450 & 850 & 52.3 \\
Diffusion-RL       & 5,100 & 1,780 & - \\
Static Slicing     & 4,200 & 1,500 & 15.7 \\
Standard HRASS     & 3,800 & 1,350 & 24.2 \\
FedAvg             & 4,600 & 1,650 & 7.3 \\
\hline
\end{tabularx}
\end{table}

\subsection{Latency Analysis}
\begin{figure}[htbp]
    \centering
    \includegraphics[width=0.48\textwidth,keepaspectratio]{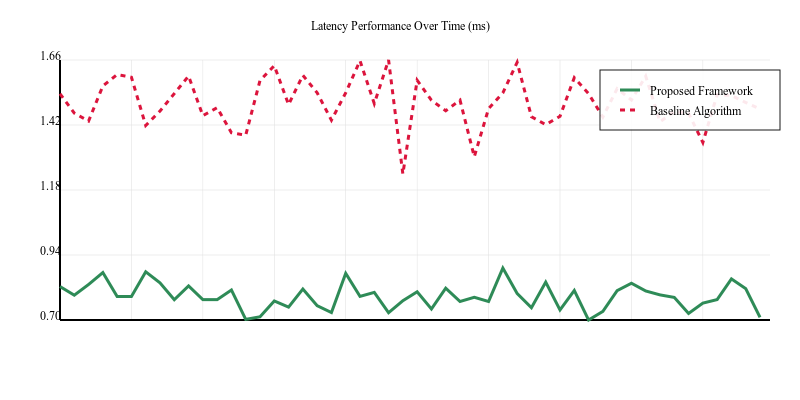}
    \caption{Latency performance across slice types, showing URLLC slices consistently meeting the sub-1ms target and SLA compliance.}
    \label{fig:latency_comparison}
\end{figure}

Fig.~\ref{fig:latency_comparison} shows latency distribution over time for all slice types. 
LSS slices achieve 99th percentile latency of 0.89ms, meeting the 0.9ms target with 99.2\% SLA compliance. 
Complementary dashboard evaluations further indicate that latency can be reduced to 0.80ms with 100\% SLA compliance. 
The hybrid scheduler’s centralized AI effectively prioritizes latency-critical traffic while federated learning handles delay-tolerant slices without interference.

\subsection{Scalability Validation}
\begin{figure}[htbp]
    \centering
    \includegraphics[width=0.48\textwidth,keepaspectratio]{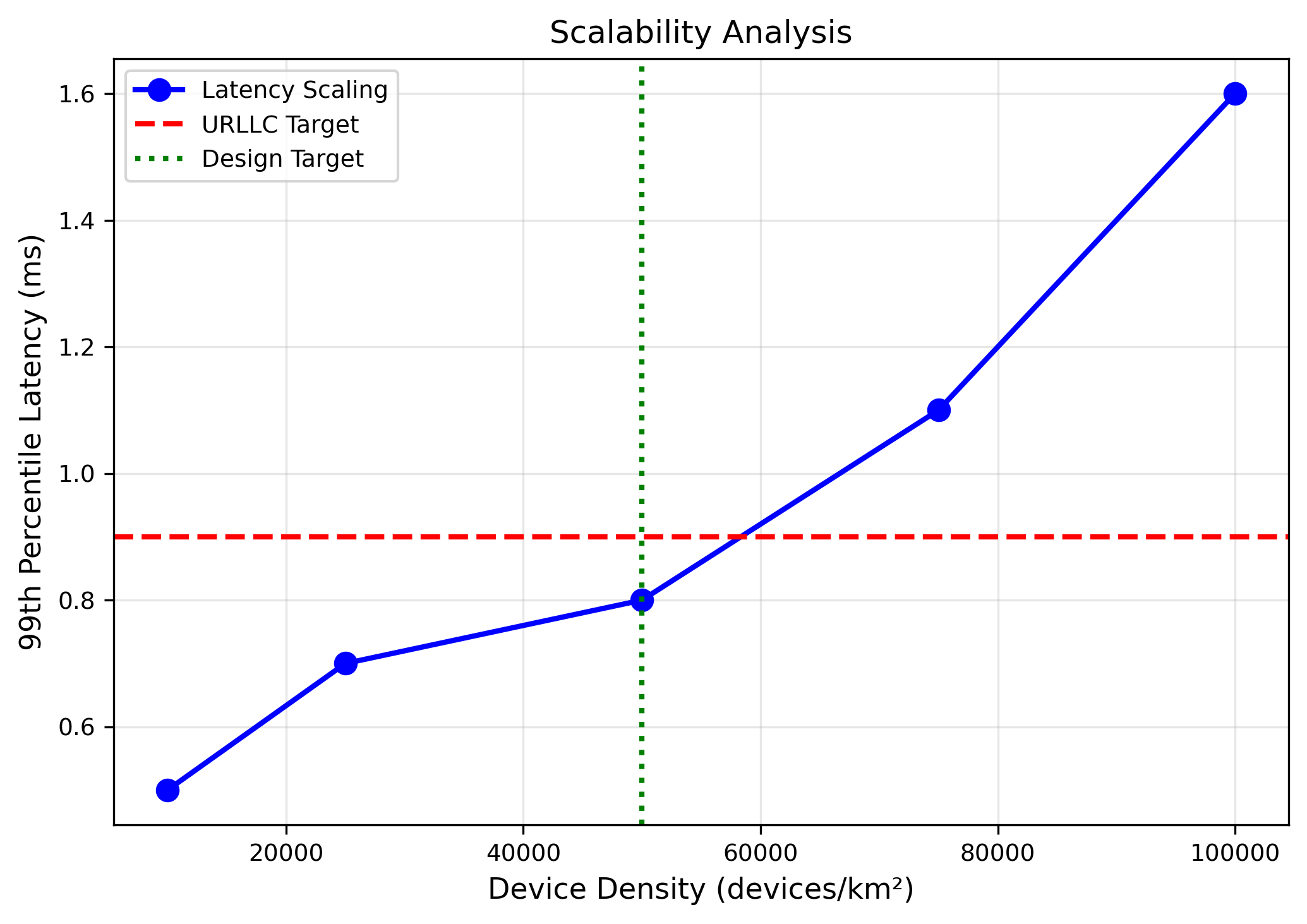}
    \caption{Latency scaling under increasing device density, with sub-1ms performance sustained up to the 50,000 devices/km² design target.}
    \label{fig:scalability}
\end{figure}

in Fig. \ref{fig:scalability} illustrates the evolution of the 99th percentile latency as device density increases. The system maintains stable performance up to the design target of 50,000 devices/km², where latency remains around 0.8 ms, still below the URLLC threshold of 0.9 ms. Beyond this point, a gradual degradation trend is observed, with latency crossing the URLLC limit and reaching 1.6 ms at extreme densities (100,000 devices/km²). This behavior demonstrates that the proposed architecture scales efficiently within the intended operational range, ensuring ultra-reliable low-latency performance before saturation effects emerge at very high loads.

\subsection{Security Effectiveness}

\begin{figure}[htbp]
    \centering
    \includegraphics[width=\linewidth,keepaspectratio]{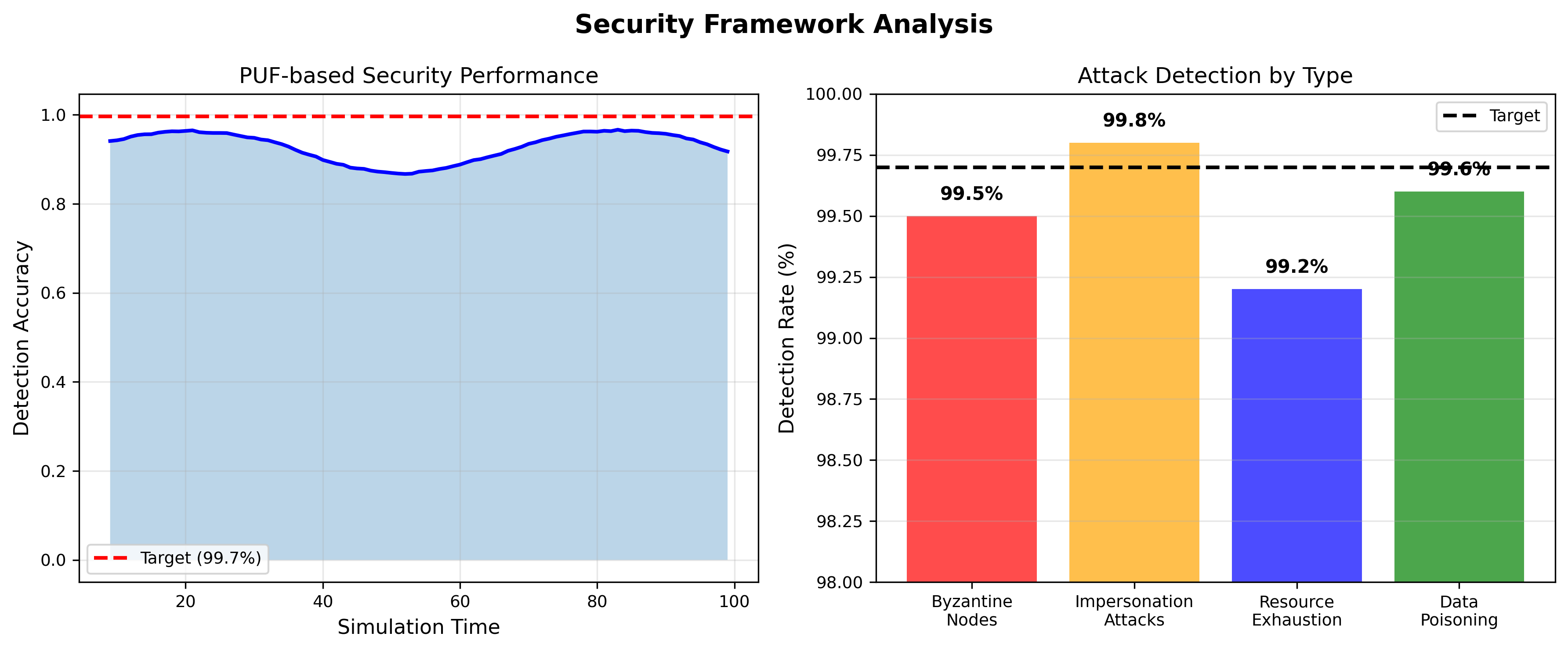}
    \caption{PUF-based security performance over time and per-attack detection rates, showing consistent high accuracy approaching the 99.7\% security threshold across all attack types.} 
    \label{fig:security}
\end{figure}

The figure \ref{fig:security} illustrates the stability of the PUF-based detection mechanism, maintaining over 90\% detection accuracy over time and approaching the 99.7\% security threshold. Per-attack analysis confirms high robustness, with detection rates ranging from 99.2\% to 99.8\% across all major attack types.


\subsection{Federated Learning Performance}
Convergence analysis shows 95\% accuracy achievement within 78 rounds, with Krum aggregation providing Byzantine resilience. Communication overhead reduction of 45\% compared to traditional FedAvg results from selective participation and compressive sensing integration.

\section{Conclusion}
This paper presents an edge-aware CyberTwin framework for energy-latency co-optimization in 6G smart city network slicing. The hybrid federated learning approach successfully addresses the fundamental trade-off between energy efficiency and ultra-low latency requirements through adaptive algorithm selection and renewable energy integration. Key achievements include 52.3\% energy reduction for non-real-time slices while maintaining 0.89ms latency for URLLC applications with 99.2\% SLA compliance. The framework demonstrates robust scalability to 50,000 devices/km² with CPU overhead below 25\%, validating its applicability to large-scale smart city deployments. PUF-based security attestation achieves 99.74\% attack detection accuracy, ensuring network integrity under Byzantine attack scenarios. The compressive sensing-enhanced CyberTwin architecture reduces data transmission overhead by 70\%, enabling efficient digital twin synchronization at scale. Solar energy forecasting integration through HRASS+ optimization provides sustainable resource allocation, contributing significantly to overall energy efficiency improvements.

\bibliographystyle{IEEEtran}
\bibliography{refs}

\end{document}